\documentclass[runningheads]{llncs}

\usepackage{graphicx}
\usepackage{amsmath}
\usepackage{amssymb}
\usepackage{hyperref}
\usepackage{color}

\usepackage{algorithm}
\usepackage{algpseudocode}
\usepackage{booktabs}
\usepackage{todonotes}
\usepackage{mathtools}

\newcommand{\expSym}{\mathbb{E}}
\newcommand{\E}[1]{\ensuremath{\expSym\left(#1\right)}}

\newcommand{\T}{\top}
\newcommand{\Var}{\ensuremath{\mathrm{Var}}}

\iftrue
\newcommand{\vw}[1]{}%{\todo[color=red]{\emph{VW}: #1}}
\newcommand{\luca}[1]{}%{\todo[color=green]{\emph{LB}: #1}}
\renewcommand{\todo}[1]{}
\else
\newcommand{\vw}[1]{\todo[color=red]{\emph{VW}: #1}}
\newcommand{\luca}[1]{\todo[color=green]{\emph{LB}: #1}}
\fi

\spnewtheorem{model}[theorem]{Model}{\bfseries}{\itshape }

\begin{document}
\title{Bounding Mean First Passage Times in\\   Population Continuous-Time
Markov Chains}
\titlerunning{Bounding Mean First Passage Times in PCTMCs}
\author{Michael Backenk\"ohler${}^{1}$, Luca Bortolussi${}^{2,1}$, Verena Wolf${}^{1}$}
\authorrunning{M.\ Backenk\"ohler et al.}
\institute{${}^{1}$Saarland University, Germany,\\ ${}^{2}$ Saarbr\"ucken Graduate School of Computer Science,\\
${}^{2}$University of Trieste, Italy}
\maketitle
\begin{abstract}
  We consider the problem of bounding mean first passage times and reachability probabilities
  for the class of population continuous-time Markov chains,  which capture stochastic interactions between groups of identical agents.
  The  quantitative analysis of such models
  is notoriously difficult since typically neither   state-based numerical
  approaches nor methods based on stochastic sampling give efficient and accurate results.
  Here, we propose a novel 
  approach that leverages techniques from martingale theory and stochastic processes to generate constraints on the statistical moments of first passage time distributions. These constraints induce a semi-definite program that can be used to compute exact bounds on reachability probabilities and mean first passage times without numerically solving the transient probability distribution of the process or sampling from it.
We showcase the method on some test examples and tailor it to  models exhibiting multimodality, a class of particularly challenging scenarios from biology.
\end{abstract}

\section{Introduction}
Population Continuous-Time Markov Chains (PCTMCs)    provide a
widely used framework to capture stochastic interactions between groups of identical agents.
This subclass of Continuous-Time Markov Chains (CTMCs)  is used
to describe the stochastic dynamics of systems in various domains.
Prominent applications are chemical reaction networks in quantitative
biology~\cite{BuchWolkenhauer},
epidemic spreading~\cite{porter2016dynamical}, performance analysis  of technical and
information systems~\cite{bortolussi2013,gast2019} as well as the behavior of
collective adaptive systems~\cite{bernardo2016}.

For the quantitative analysis of CTMCs, many   approaches have been
developed, where properties of interest are often expressed in terms of temporal logics such as
CSL~\cite{aziz1996verifying,baier2000model,baier2003model},
MTL~\cite{chen2011time}, and timed-automata
specifications~\cite{chen2009quantitative,mikeev2013fly}.
In addition, there exist
efficient software
tools~\cite{hinton2006prism,kwiatkowska2011prism,dehnert2017storm}
that can be used to analyze and verify system properties.
The computation of reachability probabilities is a central problem in this context.

Popular exact methods for CMTCs rely on numerical approaches that explicitly consider each system state individually.
A major problem is that these methods cannot scale in the context of population
models with large copy numbers of agents.
A popular alternative to tackle this problem is statistical model checking,
which is based on stochastic simulation~\cite{david2015statistical}.
For PCTMCs   arising in the context of chemical reaction networks, trajectories of the process are
usually generated using the Stochastic Simulation Algorithm
(SSA)~\cite{gillespie77}.
However, since the number of possible interactions grows with the number of
agents, stochastic simulations of PCTMCs are  time-consuming. Moreover, they are
subject to inherent statistical uncertainty and give only statistically estimated bounds.

As an alternative, recent work concentrates on numerical methods that
approximate the statistical moments of the system without the need to
compute the probability of each state.
For groups of identically behaving agents, it is possible to derive systems of
differential equations
for the evolution of the statistical population
moments~\cite{bogomolov2015adaptive,schnoerr2017survey,bortolussi2013model,engblom2006computing,schnoerr2015,gast2019}.
However, as the system of exact moment equations is infinite-dimensional, approximation
schemes typically rely on certain assumptions about the
underlying probability distribution to truncate it.
For example, one might employ a ``low dispersion closure'' which assumes that
higher-order moments are the same as those of a normal
distribution~\cite{hespanha2008moment}. %\vw{Hespanha macht doch lognormal?}\todo{in dem papier macht er eine survey f\"ur sein tool}
Such approximations are, by nature, ad-hoc and do not come with
any guarantees.

Moment-based methods often scale well   in terms of population sizes. However,
it is not possible to control the effects of the introduced approximations, which in some
cases can lead to large errors~\cite{schnoerr2015}.
This issue reverberates
on the application of these methods to  compute reachability
probabilities and mean first passage
times~\cite{hayden2012fluid,bortolussi2013model,bortolussi2014stochastic}.
Moreover, they can suffer from numerical instabilities, in particular, when the
maximum order of the considered moments has to be increased to more
appropriately describe the underlying distribution. %, a fact impacting their
%scalability.

Here, we put forward a method based solely on moments that gives \emph{exact bounds}  for Mean First Passage Times (MFPTs) and reachability probabilities in PCTMCs. For a set of states, the MFPT within a fixed
time-horizon $T$  directly characterizes the probability of reaching that set
within $T$ time units. Thus, safe upper and lower bounds on MFPTs can constitute a core component for the verification of properties in PCTMCs.
Our approach extends recent work on moment
bounds~\cite{sakurai2017convex,dowdy2018dynamic} and it is based on a martingale formulation of the stopped process that we
derive from the exact moment equations. From this formalization, we deduce a set
of linear moment constraints from which we derive
upper and lower moment bounds using semi-definite programming (SDP). Monotone
sequences of both upper and lower bounds
can be obtained by increasing the order of the relaxation. Crucially, no
closure approximations are introduced.
Therefore the bounds are exact up to the numerical accuracy of the SDP solver.

To experimentally validate our method in terms of accuracy and feasibility, we run some tests on examples from biology, leveraging an existing SDP solver and obtaining encouraging results. 
Comparing with other moment-based methods, our approach is not based on approximations due to closure schemes, thus providing guarantees on the bounds   up to the numerical accuracy of the computations. However, similarly to other moment-based methods, we also found the insurgence of numerical instabilities because moments of higher order tend to span over many orders of magnitude. We ameliorate this problem by considering scaling strategies that reduce such variability. 
We also extend our approach to deal with PCTMCs exhibiting strong multimodal behavior, due to the presence of populations having low copy numbers. This extension exploits some ideas from hybrid moment closures~\cite{kazeroonian2014modeling}.

% For our numerical investigations, we concentrate  and find encouraging
% results already for a small number of moments.
% For instance, in one of our case studies  $100,\!000$ SSA runs are necessary to
% achieve a relative   width of 0.9\% for the MFPT confidence interval.
% The SDP solver, however, returns a guaranteed interval with 
% a relative width of 0.3\%.\\

\noindent
In summary, this paper presents the following novel contributions:
\begin{itemize}
    \item the derivation of moment constraints, based on a martingale formulation, for bounding first passage times
      and reachability probabilities using a convex programming scheme;
      %using a martingale formulation;
    \item the extension of this scheme using hybrid moment conditions to systems
      exhibiting multimodal behavior;
    \item a scaling strategy for improved robustness during optimization
\end{itemize}

The paper is structured as follows:
Section~\ref{sec:related} covers work related to the analysis of first passage
times in PCTMCs and recent work on moment bounds. Section~\ref{sec:bg}
introduces the PCTMC framework and its semantics. In Section~\ref{sec:moments} we
derive a martingale from the moment dynamics of a PCTMC\@. Based on this
process, in Section~\ref{sec:mfpt_bounds} we formulate linear and semi-definite constraints
to state a semi-definite program to compute bounds on the MFPT and reachability
probabilities. In Section~\ref{sec:evaluation}, we discuss the practical
considerations of the SDP implementation and provide results on a set of case
studies. Finally, in Section~\ref{sec:conclusion} we provide concluding remarks
and directions of future work.

\section{Related Work}\label{sec:related}
Considerable effort has been directed at the analysis of first passage time
distributions in PCTMCs. Most works can either focus on an explicit state-space
analysis~\cite{barzel2008calculation,munsky2009specificity,kuntz2019exit,kuntz2018approximation}
or employ approximation techniques for which, in general, no error bounds can be
given~\cite{schnoerr2017efficient,hayden2012fluid,bortolussi2014stochastic}.
For some model classes such as kinetic proofreading, analytic solutions are
possible~\cite{munsky2009specificity,bel2009simplicity,iyer2016first}.

Barzel and Biham~\cite{barzel2008calculation} propose a recursive scheme that consists of one equation for
each state, expressing the average time the system needs to transition from that
state to the target state.
Kuntz et al.~\cite{kuntz2018approximation} propose to employ moment bounds in a
linear programming approach to compute exit time distribution using state-space
truncation schemes. In Ref.~\cite{kuntz2019exit} the authors propose a finite state-space
projection scheme to bound first passage time distributions

Hayden et al.~\cite{hayden2012fluid} use moment closure approximations and
Chebychev's inequality to gain an understanding of first passage time dynamics.
Schnoerr et al.~\cite{schnoerr2017efficient} also employ a moment closure approximation
and further approximate threshold functions to derive an approximate first passage time
distribution.
Bortolussi and
Lanciani~\cite{bortolussi2014stochastic} use a mean-field approximation
which is required to reach the target region.

Recently, several groups independently suggested the use of semi-definite
optimization for the computation of moment bounds for the limiting
distribution~\cite{ghusinga2017exact,dowdy2018bounds,kuntz2017rigorous,sakurai2017convex}.
In this approach, the differential equations describing the moment dynamics are
set to zero and form linear constraints~\cite{backenkohler2017moment}. Alongside, semi-definite constraints can
be placed on the \emph{moment matrices}. These give a semi-definite program
that can be solved efficiently.

This approach has been extended to the transient
case~\cite{dowdy2018dynamic,sakurai2019bounding}.
The approach is similar in both works and is a cornerstone of the MFPT analysis
presented here.
They differ mainly by the fact that Sakurai and Hori apply a polynomial
time-weighting~\cite{sakurai2019bounding}, while Dowdy and Barton use an
exponential one~\cite{dowdy2018dynamic}. We adopt the former approach because it
can be naturally adapted to the description of densities over time.
The resulting forms can also be adapted to statistical estimation
problems~\cite{backenkohler2019control}.

Semi-definite programming has been applied to a wide range of problems,
including stochastic processes in the context of financial
mathematics~\cite{lasserre2006pricing,kashima2009polynomial}.
For good introductions and overviews of application areas, we refer
the reader to Parrilo~\cite{parrilo2003semidefinite} and, more recently,
Lasserre~\cite{lasserre2010moments}.

Particularly relevant for this work is the application of convex optimization to
first passage times.
Helmes et al.~\cite{helmes2001computing} formulated a linear program using the
%basic adjoint equation and
Hausdorff moment conditions to bound moments of the
first passage time distribution in Markovian processes.
Semi-definite optimization has been successfully applied in financial
mathematics by Kashima and Kawai~\cite{kashima2009polynomial}, as well as
Lasserre et al.~\cite{lasserre2006pricing} to bound prices of exotic options.
%Here, the approach by Lasserre is adapted to PCTMCs.

\section{Preliminaries}\label{sec:bg}
A Population Continuous-Time Markov Chain (PCTMC)   describes the interactions
among
a set of agents of $n_S$ types $S_1,\dots, S_{n_S}$ in a well-stirred reactor. In the
sequel, we will also use other letters than $S_i$ as agent types.
Since we assume that all agents are equally distributed in space,
we only keep track of the overall copy number of agents for each type.
Therefore the state-space is $\mathcal{S}\subseteq\mathbb{N}^{n_S}$.
The  interactions are expressed as \emph{reactions} with a certain
gain and loss of agents, given by the non-negative integer vectors
$\vec{v}_j^{-}$ and $\vec{v}_j^{+}$ for
some reaction $j$, respectively. Such a reaction is denoted as
\begin{equation}\label{eq:reaction}
    \sum_{i=1}^{n_S} v_{ji}^{-} S_i
    \xrightarrow{a_j}
    \sum_{i=1}^{n_S} v_{ji}^{+} S_i\,.
\end{equation}
The reaction rate constant $a_j>0$ determines the propensity function
$\alpha_j$
of the reaction.
If just a constant is given, \emph{mass-action} propensities are assumed, where
for $\vec x\in\mathcal{S}$ we define
\begin{equation}\label{eq:stoch_mass_action}
    \alpha_j(\vec{x})\coloneqq a_j\prod_{i=1}^{n_S}\binom{x_i}{v_{ji}^{-}}\,.
\end{equation}
This choice of propensity function is natural, since it is proportional to
the number of reactant combinations.
The system's behavior is described by a stochastic
process $\{{\vec{X}}_t\}_{t\geq 0}$. % defined on the probability space $(\Omega, \mathcal{F}, \Pr)$.
We denote the abundance of a given agent type $S_i$ in $\vec X_t$ by
$X_t^{(S_i)}$.
The propensity  $\alpha_j(\vec x)$ gives the infinitesimal probability
of a reaction occurring, given a state $\vec x$. That is, for
$\vec v_j=\vec v_j^+-\vec v_j^-$ and a
small time step $\Delta t >0$,
\begin{equation}\label{eq:reaction_prob}
    \Pr({\vec{X}}_{t+\Delta t}=\vec{x}+\vec{v}_j\mid \vec{X}_t=x)
    =\alpha_j(\vec{x})\Delta t + o(\Delta t)\,.
\end{equation}
Therefore, given a system of $n_R$ reactions, the semantics of $\vec X_t$ is
given by a continuous-time
Markov chain (CTMC) on $\mathcal{S}$ with infinitesimal  generator matrix $Q$
with entries
\begin{equation}\label{eq:cme_generator}
    Q_{\vec x, \vec y} = \begin{cases}
        \sum_{j:\vec x+\vec v_j = y}\alpha_j(\vec x)\,,&\text{if}\;\vec x\neq
        \vec y,\\[1ex]
        -\sum_{j=1}^{n_R} \alpha_j(\vec x)\,, &\text{otherwise.}
    \end{cases}
\end{equation}
Accordingly,
given an initial distribution   on $\mathcal{S}$, the time-evolution of the process' distribution is given by the
Kolmogorov forward equation.
For a single state, in the context of quantitative biology, it is commonly referred to
as the \emph{chemical master equation} (CME)
\begin{equation}\label{eq:cme}
    \frac{d\pi}{d t} (\vec x,t) =
    \sum_{j=1}^{n_R}\left(
        \alpha_j(\vec x-\vec v_j)\pi(\vec x-\vec v_j,t) - \alpha_j(\vec x)\pi(\vec x,t)
    \right)\,,
\end{equation}
where $\pi(\vec x,t)=\Pr(\vec X_t=\vec x)$ and $\Pr(\vec X_0=\vec x)=\pi(\vec{x}, 0)$.

\noindent
Consider the following simple PCTMC with non-linear propensities as an example.
\begin{model}[Dimerization]\label{model:dim}
  We first examine a simple dimerization model on an unbounded state-space with
  reactions
  $$\varnothing\xrightarrow{\lambda}M,\quad 2M\xrightarrow{\delta}D$$
  and initial condition $X_0^{(M)}=X_0^{(D)}=0$. The semantics is given by a
  CTMC
  $\vec{X}_t=(X_t^{(M)}, X_t^{(D)})^{\T}$, where $(S_1, S_2)=(M,D)$. The reaction
  propensities according to~\eqref{eq:stoch_mass_action}
  are $\alpha_1(\vec{x})=\lambda$ and $\alpha_2(\vec{x})=\delta\, x^{(M)} (x^{(M)} -
  1)/2$. The change vectors $v_1^-={(0,0)}^{\T}$, $v_1^+={(1,0)}^{\T}$,
  $v_2^-={(2,0)}^{\T}$, and $v_2^{+}={(0,1)}^{\T}$.
  Consequently, $v_1={(1,0)}^{\T}$ and $v_2={(-2, 1)}^{\T}$.
\end{model}
For a state $(x^{(M)}, x^{(D)})\in\mathbb{N}^2$, where $x^{(M)}\geq 2$,
the CME \eqref{eq:cme} becomes
\begin{align*}
\frac{d}{dt}\pi((x^{(M)}, x^{(D)}), t)
=&\lambda \pi((x^{(M)}-1, x^{(D)}), t)\\ &+ \frac{\delta}{2} \, (x^{(M)}+2) (x^{(M)} +
  1) \pi((x^{(M)}+2, x^{(D)}-1), t)\\ &- (\lambda + \frac{\delta}{2} \, x^{(M)} (x^{(M)} -
  1))\pi((x^{(M)}, x^{(D)}), t)\,.
\end{align*}
This explicit representation of state probabilities is often not possible, because
there are infinitely many states. Usually the state-space is truncated to contain all
relevant states~\cite{andreychenko2011parameter} or one switches to
an approximation such as the mean-field~\cite{bortolussi2013}.

In this work, we are interested in \emph{first passage times} of such processes.
That is the time, the process first enters a set of target states
$B\subseteq \mathcal{S}$. Naturally, the analysis of first passage times is
equivalent
to the analysis of times at which the process exits the complement $\mathcal{S}\setminus B$.
More formally, the first passage time $\tau$ for
some target set $B$ is defined as the random variable
\begin{equation}\label{eq:fpt_def}
    \tau = \inf\{t\geq 0\mid \vec X_t \in B\}\,.
\end{equation}

In this example, we are interested in the time at which the number of type $M$ agents
exceed some threshold $H$.
With the framework presented in the sequel, one can bound the expected value
of this time using semi-definite programming.
Further, it is possible to impose a time-horizon $T$, and find bounds
on the probability of $ X_t^{(M)}\geq H$ for some $0\leq t\leq T$.
The employed framework is centered around semi-definite relaxations
of the generalized moment problem~\cite{lasserre2010moments}.
These require linear constraints on the moments of measures.
In the following section, we derive such constraints.

\section{Martingale Formulation}\label{sec:moments}
Next, we will discuss the ordinary differential equations for the evolution of the statistical moments of
the process.
The moments over the state-space are then used to derive temporal moments, i.e.\ moments
of measures over both the state-space and the time.
This extended description results in a process with the martingale property.
This property can be used to formulate linear constraints on the temporal moments
and, as a special case, the mean first-passage time.
In combination with semi-definite properties of moment matrices, we can formulate
mathematical programs that yield upper and lower bounds on mean first passage times.

We start with the description of the \emph{raw moments} dynamics.
In particular, a raw moment is
$$\E{\vec X^{\vec m}}=\E{\prod_{i=1}^{n_S} X_i^{m_i}}\,, \quad\vec m\in \mathbb N^{n_S}$$ with respect to some probability measure.
%Note, that moments can also be defined for non-probability measures, i.e.\ for
%measures with total measure $\neq 1$.
The order of a moment $\E{{\vec X}^{\vec m}}$ is given by the sum of its exponents,
i.e.\ $\sum_i m_i$.
Note that the notion of  expected value can be generalized
to any measure $\mu$ on a Borel-measurable space
$(E, \mathcal{B}(E))$, where
 the $\vec{m}$-th raw moment is $\int_E {\vec x}^{\vec m}\,d\mu(\vec x)$.
Throughout we assume that moments of arbitrary order remain finite over time,
i.e.\ $\E{\lvert \vec{X}^{\vec{m}}\rvert}<\infty$, $t\geq 0$.
In Ref.~\cite{gupta2014scalable} the authors propose a framework to verify
this property for a given model.

Let $f$ be a polynomial function, $t\ge0$.
Using the CME \eqref{eq:cme}, we can derive ordinary differential equations (ODEs)
describing the dynamics of $\E{f(\vec{X}_t)}$~\cite{engblom2006computing}.
Specifically,
\begin{equation}\label{eq:mom_ode}
    \frac{d}{dt}\E{f(\vec X_t)} = \sum_{j=1}^{n_R}\E{\left(f({\vec X_t +
    \vec{v_j}}) - f(\vec X_t)\right)\alpha_j(\vec X_t)}\,.
\end{equation}

Let us consider Model~\ref{model:dim} as an example and
agent type $M$. Further, let $X_t=X_t^{(M)}$
for ease of exposition.
When choosing $f(X_t)=X_t^m$, $m=1$ and   $m=2$
we obtain two differential equations describing
the change of the first two moments of species $M$,
$\E{X_t}$ and $\E{X_t^2}$, respectively.
\begin{align}
    \frac{d}{dt}\E{{X}_t} &= \lambda\E{{X}_t^0} -
    2{\delta}\left(\E{{X}_t^2}-\E{{X}_t}\right)\label{eq:dim_exp_ode}\\[1ex]
    \frac{d}{dt}\E{{X}_t^2} &= \lambda(2\E{{X}_t} + 1) - 4\delta\left(\E{{X}_t^3} -
    2\E{{X}_t^2} + \E{{X}_t}\right)\,.
\end{align}
Fixing initial moments, the ODE system describes the moments over time exactly.
However, these ODEs cannot be integrated because the system is not closed.
The right-hand side for moment $\E{X_t^m}$ always contains $\E{X_t^{m+1}}$.
To solve the initial value problem,
one typically resorts to ad-hoc approximations of the highest order moments
to close the system. Here we do \emph{not} need such approximations
because we do not numerically integrate the moment equations.
Instead we adopt an approach \cite{dowdy2018dynamic,sakurai2019bounding} that extends
the description of state-space moments to a temporal one.

This is achieved by the introduction of a time-dependent polynomial $w(t)$ that is multiplied to
\eqref{eq:mom_ode}.
An integration by parts on $[0, T]$ yields~\cite{dowdy2018dynamic,sakurai2019bounding}
\begin{equation}\label{eq:exp_constraint}
\begin{split}
        & w(T)\,\E{f(\vec X_T)}
        - w(0)\,\E{f(\vec X_{0})}
        - \int_{0}^{T}\frac{dw(t)}{dt}\E{f(\vec X_t)}\,dt\\
        =&\sum_{j=1}^{n_R}\int_{0}^{T}w(t)\,
        \E{\left(f{(\vec X_t + \vec v_j)} - f(\vec X_t)\right)\alpha_j(\vec X_t)}\,dt\,.
        \end{split}
\end{equation}
We now want to interchange the order of integration and the summation due to the expected value.
To this end, we have to assume the absolute convergence of the integrals.
On finite time intervals $[0,T]$ this holds because $w$
is polynomial and we assumed finite moments for all $t\geq 0$.
Interchanging the  summation  and integral of a monomial
${\vec{x}}^{\vec{m}}$, i.e.\ pulling all expectation operators outside
\begin{equation*}
\begin{split}
\int_{0}^Tg(t)\E{\vec X_t^{\vec m}}\,dt
% =&\int_{t_0}^t\sum_{x\in\mathcal{S}}g(s)\Pr(X_s=x)x^m\,ds\\
% =&\int_{t_0}^t\int_{\Omega} g(s) {X_s(\omega)}^m\,dP(\omega)\,ds\\
% =&\int_{\Omega}\int_{t_0}^t g(s) {X_s(\omega)}^m\,ds\,dP(\omega)\\
=&\;\E{\int_{0}^Tg(t){\vec X}_t^{\vec m}\,dt}.
\end{split}
\end{equation*}
Hence, we are able to to pull out the expectation operator in~\eqref{eq:exp_constraint}.
\begin{equation}\label{eq:e_exp_constraint}
\begin{split}
    0=&\,w(T)\E{f(\vec X_T)} - w(0)\E{f(\vec X_{0})} -
    \E{\int_{0}^T\frac{dw(t)}{dt}f(\vec X_t)\,dt}\\
    &-\sum_{j=1}^{n_R}\E{\int_{0}^Tw(t)
         (f(\vec X_t+\vec v_j) - f(\vec X_t))\alpha_j(\vec X_t)\,dt}\,,
         \end{split}
\end{equation}
This gives us the expected value of a time-dependent function of the original process.
The function can be viewed as a stochastic process of its own where the
time-horizon $T$ is the index variable. A key property
of this process is also illustrated by \eqref{eq:e_exp_constraint}: The process'
expected value remains 0, regardless of the choice of $T$.
This martingale property is particularly useful because it can be used
to formulate linear constraints on stopping times of the process.
Explicitly, we can define this process $\{Z_T\}_{T\geq 0}$ parameterized by
the time-weighting $w$ and polynomial $f$.
\begin{equation}\label{eq:martingale}
\begin{split}
    Z_T\coloneqq&\,w(T)f(\vec X_T) - w(0)f(\vec X_{0}) -
    \int_{0}^T\frac{dw(t)}{dt}f(\vec X_t)\,dt\\
    &-\sum_{j=1}^{n_R}\int_{0}^Tw(t)
         (f(\vec X_t+\vec v_j) - f(\vec X_t))\alpha_j(\vec X_t)\,dt\,,
         \end{split}
\end{equation}
A useful choice for $f$ and $w$ are monomials.
When choosing $w(t)=t^k$ with $k\in\mathbb N$ and $f(\vec X)={\vec X}^{\vec m}$
the process takes the form
\begin{equation}\label{eq:basic_poly_martingale}
Z_T^{(\vec m, k)}=
         T^k \vec X_T^{\vec m}
        - 0^k \vec X_{0}^{\vec m}
        % - k \int_0^T t^{k-1} \vec X_t^{\vec m}\,dt
        + \sum_{i}c_i\int_0^T t^{k_i} \vec X_t^{\vec m_i}\,dt
\end{equation}
where   $(\vec m_i)_i$, $(k_i)_i$, and $(c_i)_i$ are finite sequences resulting
from the substitution
of $f$ and $w$
and expansion of~\eqref{eq:martingale}.
 This choice allows to naturally
characterize the behavior in time and state-space as moments, because
the expected value of \eqref{eq:basic_poly_martingale} then becomes a linear form
of moments.
We will use these as constraints in the semi-definite program used to bound MFPTs.

% \todo{re-state the above example here again?}\luca{good idea}
If we apply this to our previous example~\eqref{eq:dim_exp_ode}, letting $m=1$ %\vw{same example has $\vec m=(1,0)$ above }
and $k=1$ we obtain the following process for Model~\ref{model:dim}.
\begin{align*}
    Z_T^{(1,1)} = TX_T - \int_0^T X_t\,dt - \lambda \int_0^T t\,dt - 2\delta
    \int_0^T t X_t\,dt +
    2{\delta}\int_0^TtX_t^2\,dt,
\end{align*}
where the sequences above are $(m_i)_i=(1,0,1,2)$, $(k_i)_i=(0,1,1,1)$,
and $(c_i)_i=(-1,-\lambda, -2\delta,2\delta)$.
\section{Bounds for Mean First Passage Times}\label{sec:mfpt_bounds}
We now turn to the analysis of first passage times within some time-bound
$T>0$. Given some subset of the state-space
$B\subseteq \mathcal{S}$ the first passage time is given by the continuous random variable
\begin{equation}
    \tau=\inf\{t\geq 0\mid \vec X_t \in B\}\land T\,,
\end{equation}
where $a \land b \coloneqq \min\{a, b\}$.
For this work, we only look at threshold hitting times,
i.e.\ we set a threshold $H$ for species $S$ and thus $B=\{\vec{x}\mid x^{(S)}\geq
H\}$. Note, that this framework
allows for a more general class of target sets, which are discussed in
Section~\ref{subsec:multidim}.
In the sequel, we will use $\tau$ as a stopping time in our martingale
formulation and consider
$Z_\tau^{(\vec m, k)}$ instead of $Z_T^{(\vec m, k)}$.
Since~\eqref{eq:basic_poly_martingale} defines a martingale, $Z_{\tau}^{(\vec m, k)}$
remains a martingale by
Doob's optional sampling theorem~\cite{gihmantheory}. In particular, this
implies that $\expSym(Z_{\tau}^{(\vec m, k)})=0$ for all moment orders $m$ and
degrees $k$ in the weighting function $w(t)$.

\subsection{Linear Moment Constraints}
To simplify our presentation, we fix an initial state $\vec x_0$, i.e.\ $P(\vec X_0=\vec x_0)=1$.
Using $\expSym(Z_{\tau}^{(\vec m, k)})=0$ and the form
\eqref{eq:basic_poly_martingale} for $Z_{\tau}^{(\vec m, k)}$
yields the following linear constraint on expected values.
\begin{equation}\label{eq:constraint}
    0 = \,\E{{\tau}^k\vec X_{\tau}^{\vec m}}
        - 0^k\vec x_0^{\vec m}
        % - k \E{\int_{0}^{\tau} t^{k-1} \vec X_t^m\,dt}
        + \sum_{i}c_i\E{\int_{0}^{\tau} t^{k_i} \vec X_t^{\vec m_i}\,dt}\,,
\end{equation}
where $0^0=1$.
Hence, we have established a relationship between the process dynamics
up to the hitting time via expected values of the time-integrals and the final process state at
the hitting time via $\E{\tau^k {X}_{\tau}^{{m}}}$.

For the ease of exposition, we now turn to the analysis of first passage times in
one-dimensional processes w.r.t.\ an upper threshold $H$. In particular,
we will consider  moments $X^m$ of a one-dimensional process for $m=0,1,2\ldots$.
The   approach proposed in the sequel, however,
can be extended to multi-dimensional processes and more complex target sets $B$.

\begin{figure}[t]
    \centering
    \includegraphics[scale=.6]{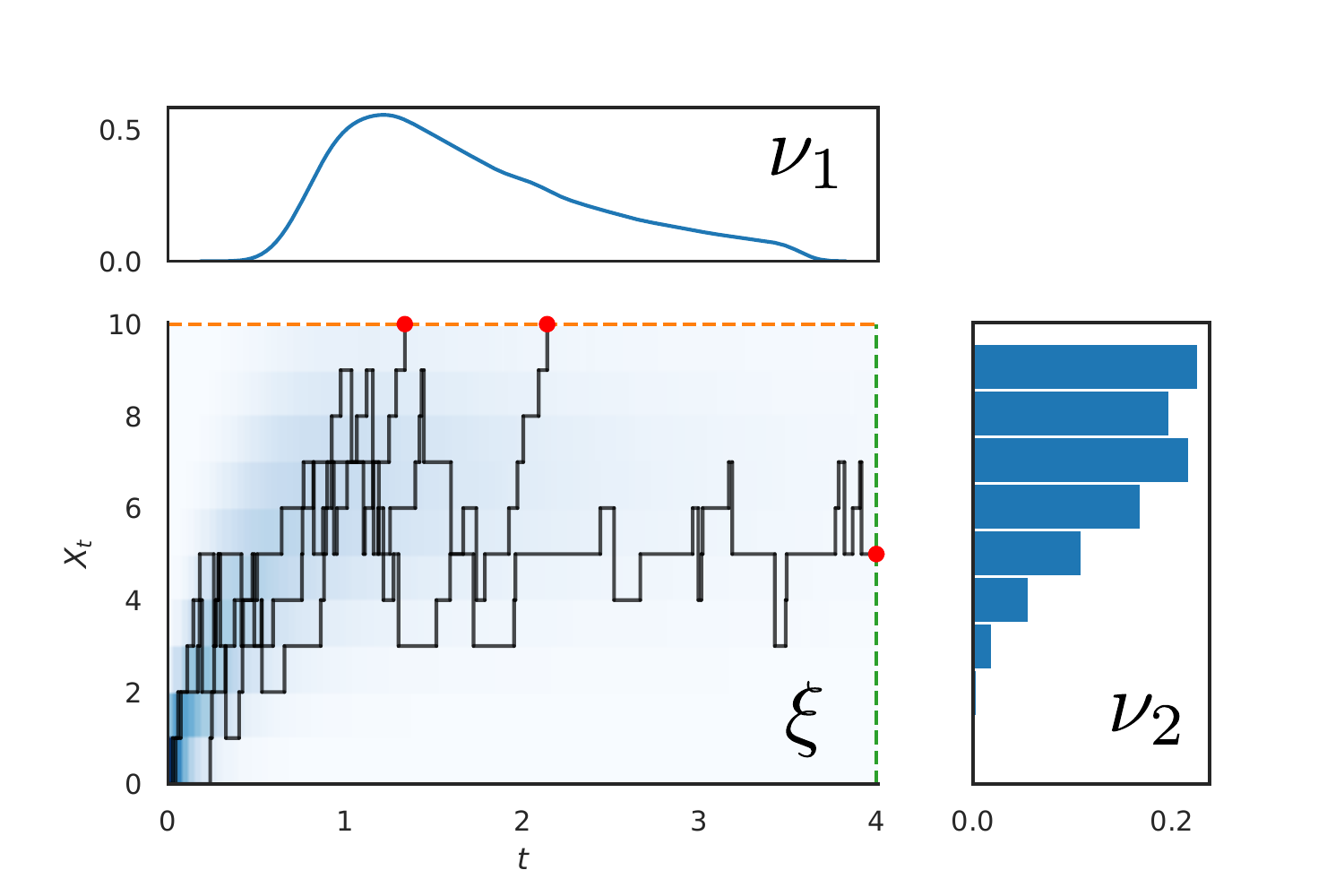}
    \caption{The relationship between the occupation measure $\xi$ and the
    exit location probability measures $\nu_1$ and $\nu_2$. The shaded area indicates
    the structure of the occupation measure. Three example trajectories are
    additionally plotted with
    their exit location highlighted. The plots are based on $10,\!000$ sample trajectories.}
    \label{fig:decomposition}
\end{figure}

Consider again Model~\ref{model:dim} and assume that we are interested in the
time at which species $M$ exceeds  threshold $H$ while fixing the considered time-horizon to
$T=4$. That is, we are interested in the stopping time $\tau=\inf\{t\geq 0\mid X_t\geq 10\}\land 4$.
Since the abundance of $D$ does not influence $M$, we can ignore
species $D$ and treat the process as one-dimensional.
Figure~\ref{fig:decomposition} shows three example trajectories:
Two reach an upper threshold $H=10$, while one reaches the final time-horizon $T=4$
The figure also illustrates another aspect present in \eqref{eq:constraint}.
It gives a connection between the terminal distribution, i.e. the distribution of $X_{\tau}$,
and the dynamic behavior up to $\tau$.
The statistics at $\tau$ are described by a distribution whose %\luca{what are nu 1 and 2? they have not introduced before}
moments are represented by the $\E{\tau^k{\vec{X}_{\tau}}^{\vec{m}}}$ term in \eqref{eq:constraint}.
This distribution corresponding two moments encompasses both cases of how
$\tau$ can be reached. In the first case threshold $H$ is reached and the second case the process reaches the time-horizon $T$.
In the following we will define the interplay between these measures more formally.

Therefore we can view \eqref{eq:constraint} as the description of a relationship between
two measures~\cite[Chapter~9.2]{lasserre2010moments}:
\begin{itemize}
\item \emph{Expected Occupation Measure} $\xi$ supported on $[0,H]\times [0,T]$:
\begin{equation}\label{eq:ex_occ_measure}
    \xi(A\times C) \coloneqq \E{\int_{[0,\tau]\cap {C}}{1}_{\in A}(X_t)\,dt},
\end{equation}
\item \emph{Exit Location Probability} supported on $(\{H\}\times[0,T]) \cup
([0,H]\times\{T\})$:
% \todo{this can also be dropped, ie.\ we don't set a time-horizon. should be fine if passage time moments are finite (also
% in the relaxation)}
\begin{equation}\label{eq:exit_loc_measure}
    \nu(A\times C)\coloneqq \Pr((X_{\tau},\tau)\in A\times C),
\end{equation}
\end{itemize}
where $A\times C$ is a measurable set, i.e.\ $A$ and $C$ are elements of the Borel $\sigma$-algebras on $[0,H]$ and $[0,T]$, respectively.

Using Figure~\ref{fig:decomposition}, one can gain an intuition for these two measures.
The expected occupation measure is shaded in blue.
As the name implies $\xi(A\times C)$ tells us how much time  the process spends
in $A$ up
to $\tau$ restricting to the time instants belonging to $C$.
In particular, $\xi([0,H]\times [0,T])=\E{\tau}$.
The exit location probability $\nu$, while being a two-dimensional distribution, can be viewed as a composition of a density describing the time at which the process reaches $H$ (if it does) and a probability mass function on the states of the process if the time-horizon is reached without exceeding $H$.
We partition the measure $\nu$ into $\nu_1$ and $\nu_2$ by conditioning on $\tau=T$.
Thus, $$\nu_1(C)\coloneqq\Pr(\tau\in C, \tau<T)\quad\text{and}\quad\nu_2(A)\coloneqq\Pr(X_T\in
A, \tau=T)$$
and hence $\nu(A\times C)=\nu_1(C)+\nu_2(A)$.
To refer to the moments of these measures, we define \emph{partial moments}
$$
    \E{g({X}); f({Y}) = y}\coloneqq
    \E{g({X})\mid f({Y})=y}\Pr(f({Y})=y)\,,
$$
for some polynomial $g$ and some indicator function $f$. Then
$$\E{\tau^k X_{\tau}^m}=T^k\E{X_{\tau}^m;\tau=T} + H^m\E{{\tau}^k;\tau < T, X_{\tau}=H}\,.$$
The partial expectations in terms of $\nu_1$, $\nu_2$
$$
\E{X_{\tau}^m;\tau=T} + \E{{\tau}^k;\tau < T, X_{\tau}=H}
$$
Therefore the linear moment constraints have the form
\begin{equation}
    \begin{split}
        0 = \,&T^k\E{X_{\tau}^m;\tau=T} + H^m\E{{\tau}^k;\tau < T, X_{\tau}=H}\\
        &- 0^kx_0^{m}
        % - k \E{\int_{0}^{\tau} t^{k-1} \vec X_t^m\,dt}
        + \sum_{i}c_i\E{\int_{0}^{\tau} t^{k_i}X_t^{m_i}\,dt}\,.
    \end{split}
\end{equation}
Next, we consider infinite sequences of partial moments 
 $\vec{y}_1=(y_{1k})_{k\in\mathbb{N}}$, $\vec{y}_2=(y_{2m})_{m\in\mathbb{N}}$, and $\vec{z}=(z_{mk})_{(m,k)^{\T}\in\mathbb{N}^2}$
 of $\nu_1$, $\nu_2$, and $\xi$, respectively.
$$y_{1k} \coloneqq \E{{\tau}^k;\tau < T},\quad y_{2m}\coloneqq\E{X_{\tau}^m;\tau=T},\quad
z_{km}\coloneqq\E{\int_0^{\tau}t^k X_t^m\,dt}\,$$

\subsection{Objective}
Given the above measures and their corresponding moments, we can
now identify the moments we are particularly interested in.
We formulate an optimization problem with variables corresponding 
to the moments defined above.
The MFPT is exactly the zeroth moment of $\xi$,
$$z_{00}=\E{\int_0^{\tau}1_{\leq H}(X_t)\,dt} = \E{\tau}\,.$$
Therefore $z_{00}$ corresponds to the objective of the optimization problem
that gives bounds for the MFPT.
Furthermore, we can easily change the objective to the  
zeroth moment of $\nu_1$,
$$y_{10} = \E{\tau^0;\tau<T}=\Pr(\tau < T)\,.$$
This moment is the probability of reaching
threshold $H$ before reaching time-horizon $T$. Since the target set can be more complex, this formulation can be used to perform model checking on a
wide variety of properties.

Moreover, it is possible to formulate objectives not directly corresponding to
a raw moment such as the variance~\cite{sakurai2019bounding,dowdy2018bounds}.

\subsection{Semi-Definite Constraints}
The linear constraints alone are not sufficient to identify moment bounds.
We further leverage the fact that a necessary condition for a positive measure that the \emph{moment matrices}
are positive semi-definite.
A matrix $M\in\mathbb{R}^{n\times n}$ is positive semi-definite, denoted by  $M\succeq 0$ if and only if
$$ {\vec v}^T M{\vec v} \geq 0\quad \forall \vec v\in\mathbb{R}^n\,.$$
As an example, let us consider a one-dimensional random variable $Z$
with moment sequence $\vec z$. \luca{has the notion of moment  seq been defined?}
For  moment order $r$,
the entries of the $(r+1)\times (r+1)$ moment matrix $M_r(\vec x)$ are given by
the raw moments.
In particular, $$(M_r({\vec{z}}))_{ij}=z_{i+j-2}=\E{Z^{i+j-2}}$$ for
$i,j\in\mathbb{N}_r$ where $\mathbb{N}_r=\{0,1,\dots,r\}$ and the maximum order in the matrix is $2r$.
For instance,
\begin{equation}
\label{eq:m1_dim}
    M_1(\vec x) =
    \begin{bmatrix}
    x_0 & x_1 \\
    x_1 & x_2
    \end{bmatrix}
\end{equation}
needs to be positive semi-definite. By Sylvester's criterion this means $\det
M_1\geq 0$ and $x_0\geq 0$.
We can easily see that in this case this entails
$$
\det M_1=x_0x_2 - x_1^2=\E{X^2}-\E{X}^2
=\Var({X})
\geq 0\,.
$$
This restriction is natural since the variance is always non-negative.
% Even though the measures $\nu_1$, $\nu_2$, and $\xi$ are not probability measures
% this restriction remains valid for them. Either they can be normalized to
% form a probability measure, or if their mass is zero % we don't care...
This gives us the following restrictions on the moment matrices.
\begin{equation}\label{eq:sd_constraints}
M_r(\vec{z})\succeq 0, \quad M_r(\vec{y_1})\succeq 0,\quad\text{and}\quad M_r(\vec{y_2})\succeq 0
\end{equation}
for arbitrary orders $r$, providing a first tranche of moment constraints.

Furthermore, we need to enforce the restriction of the measures $\xi$, $\nu_1$, and $\nu_2$
to their supports.
This can be done, by defining non-negative polynomials
on the intended support of the measure.
For example, $\nu_2$ has support $[0,H]$. We can now define
$$
u_H(t,x) = Hx - x^2, \quad x\in \mathbb R
$$
as a polynomial that is non-negative on $[0,H]$.
Using such polynomials, we can construct \emph{localizing matrices},
which have to be positive semi-definite~\cite{lasserre2010moments}.
Applying $u_H$ to the moment matrix of measure $\nu_2$, i.e.\ $M_1(\vec{y}_2)$
$$
M_1(u_H, \vec{y_2})=
\begin{bmatrix}
    Hy_{20} - y_{22} & Hy_{21} - y_{23} \\
    Hy_{21} - y_{23} & Hy_{22} - y_{24}
\end{bmatrix}
$$
with the constraint $M_1(u_H, \vec{y_2})\succeq 0$, where the application of
a polynomial such as $u_H$ to a moment matrix
is formally defined for the multidimensional case in Section~\ref{subsec:multidim}.
Similarly, let $u_T(t, x) = Tt-t^2$ to restrict $\nu_1$ to $[0,T)$.
The expected occupation measure $\xi$ is constrained similarly to its domain
$[0,H]\times[0,T]$.
This gives us the following restrictions on the moment matrices.
\begin{equation}\label{eq:localizing_sd_constraints}
M_r(u_T,\vec{z})\succeq 0, \quad M_r(u_H,\vec{z})\succeq 0,  \quad M_r(u_T,\vec{y_1})\succeq 0,\quad M_r(u_H,\vec{y_2})\succeq 0\,.
\end{equation}

\subsection{A Semi-definite Program to Bound MFPTs}
With the linear constraints given in \eqref{eq:constraint}
and the semi-definite constraints \eqref{eq:sd_constraints} and \eqref{eq:localizing_sd_constraints} 
discussed in the previous sections, we can now formulate a
semi-definite program (SDP).
An SDP is a convex optimization problem over the set of positive semi-definite $n \times n$-matrices
$\mathcal{X}$ under linear constraints:
\begin{equation}
    \label{eq:sdp_canonical}
    \begin{split}
        \min_{X\in\mathcal{X}} \hspace{1em} & \sum_{i,j} A_{ij}^{(0)}X_{ij} \\
        \text{such that} \hspace{1em}
                 & X\succeq 0\\
            & \sum_{i,j} A_{ij}^{(k)}X_{ij} \leq b_k, \quad k=1,\dots,m
    \end{split}
\end{equation}
with constant matrices $A^{(i)}\in \mathbb{R}^{n\times n}$, $i=0,\dots,m$ and
constants $b_k\in\mathbb{R}$, $k=1,\dots,m$ to define a set of $m$ linear constraints.
Such a problem is convex and can be solved efficiently~\cite{vandenberghe2010cvxopt}.

Now we can state the SDP relaxation to the MFPT problem for any  order $0<r<\infty$.
With each moment sequence $\vec x$ we associate a sequence proxy variables $\vec{x'}$
used in the optimization problem.
\begin{equation}\label{eq:sdp_for_fpt}
    \begin{split}
        \min / \max \hspace{1em}&  z_{00}^{\prime} \\
        \text{such that}\hspace{1em} & M_r(\vec{z'})\succeq 0,
                   M_r({u}_T, \vec{z'})\succeq 0, M_r({u}_H, \vec{z'})\succeq 0\\
        & M_r(\vec{y_1'}) \succeq 0, M_r({u}_T,\vec{y_1'}) \succeq 0\\
        & M_r(\vec{y_2'}) \succeq 0, M_r({u}_H, \vec{y_2'}) \succeq 0\\
        & 0= y_{1k}' H^m -  y_{2m}'T^k - 0^k x_0^m +\sum_i c_i  z_{k_i m_i}', \quad\forall m, k
    \end{split}
\end{equation}
This SDP can be compiled to the standard form \eqref{eq:sdp_canonical}.
To this end, the moment matrices can be arranged in a block-diagonal form and the
localizing constraints \eqref{eq:localizing_sd_constraints} can be encoded
by the introduction of new variables and appropriate equality constraints.
This transformation can be done automatically using modeling frameworks
such as CVXPY~\cite{cvxpy}. We therefore only give the SDP in the more intuitive format.
This problem can be solved using off-the-shelf SDP solvers such as MOSEK~\cite{mosek},
CVXOPT~\cite{vandenberghe2010cvxopt}, or SCS~\cite{scs}.

In principle, we can choose an arbitrarily large order $r$ for the moment matrices
and their corresponding constraints, because there are
infinitely many moments.
In practice, however, the order is bounded by practical issues such as the program size
(number of constraints and variables) and numerical issues.
These issues are discussed in Section~\ref{sec:evaluation} in more detail.
Choosing a finite $r$ is a relaxation of the problem since it removes constraints regarding
higher-order moments.

\subsection{Multi-Dimensional Generalization}
\label{subsec:multidim}
For a general multi-dimensional moment sequence
$\vec{y}={(\E{\vec X^{\vec{m}}})}_{\vec{m}\in\mathbb{N}^{n_s}}$, the moment
matrix is~\cite{lasserre2010moments}
% \todo{explain the order $r$ in some detail (max order is $2r$), define
% $\mathbb{N}_r^n$}
$$M_r(\vec y)(\vec\alpha,\vec\beta)
=y_{\vec\alpha + \vec\beta},\quad\forall\vec{\alpha},
\vec{\beta}\in\mathbb{N}_r^n$$
where row and column indices, $\vec{\alpha}$ and $\vec\beta$, are ordered according to the canonical basis
\begin{equation}\label{eq:canoncial_basis}
\vec{v}_r(\vec{x}) =
{(1,x_1,x_2,\dots,x_n,x_1^2,x_1x_2,\dots ,x_1x_n,\dots ,x_1^r,\dots
	,x_n^r)}^T\,.
\end{equation}
Equivalently, $M_r(\vec{y})=\E{\vec{v}_r (\vec x)\vec{v}_r(\vec x)^T}$.
For a moment sequence the semi-definite restriction $M_r(\vec{y})\succeq 0$ must hold.

Measures can be restricted to semi-algebraic sets
$\{\vec x\in\mathbb{R}^n \mid u_j(\vec x)\geq 0, j=1,\dots,m\}$,
where $u_j$, $j=1,\dots,m$ are polynomials~\cite{lasserre2010moments}.
This is done by placing restrictions on the {localizing matrices}.
For each polynomial $u_i\in\mathbb{R}[x]$ with coefficient vector
$\vec{u}=\{u_{\vec\gamma}\}$,
i.e.\ $u(\vec x) = \sum_{\vec{\gamma}\in\mathbb{N}^n} u_{\vec{\gamma}}
\vec{x}^{\vec{\gamma}}$,
the localizing matrix is
$$ M_r(u, \vec{y})(\vec{\alpha}, \vec{\beta})=
\sum_{\vec\gamma\in\mathbb{N}^n}u_{\vec\gamma}
y_{\vec\gamma+\vec\alpha+\vec\beta},\quad
\forall\vec{\alpha},\vec{\beta}\in\mathbb{N}^n_r.$$
Requiring that this matrix is positive semi-definite restricts the measure to
$\{\vec{x}\mid u_i(\vec{x})\geq 0\}$.
This way we can, for example, restrict the moment sequence $\vec{y}$
to measures that are positive w.r.t.\ dimension $j$.
Simply letting $u(\vec{x}) = x_j$ and requiring
$M_1(\vec{u},\vec{y})\succeq 0$ for $i=1,\dots,n_S$ gives us this restriction.

% \luca{I think here we need a subsection discussing how to compute the reachability probabilities in this framework, rather than burying it in the example sectio speces} % introduced a new section

\section{Implementation and Evaluation}\label{sec:evaluation}
The implementation of the SDP \eqref{eq:sdp_for_fpt} is straightforward using
modeling frameworks and off-the-shelf solvers.
However, as noted in previous work~\cite{dowdy2018dynamic,sakurai2017convex,dowdy2018bounds,sakurai2019bounding} on moment-based SDPs
the direct implementation of the problem may lead to difficulties for the solver.
A source of these is that moments of various orders by nature
may differ by many orders of magnitude.
A re-scaling of the moments~\cite{dowdy2018bounds,sakurai2019bounding}
such that moments only vary by few orders of magnitude
may alleviate this problem.
In other scenarios such as the bounding of general transient or steady-state moments,
the scaling can be particularly difficult,
because the magnitude of moments is generally not known
a priori. However, for the MFPT problem, we propose the following moment scaling.

\subsection{Moment Scaling}
Using the fact that
$\mathcal{S}\setminus {B}$ is often finite,
it is possible to derive trivial bounds, which can be used to scale moments.
If, for example, we have a one-dimensional process $X_t$ with $X_0 = 0$ a.s.\ and are interested in the hitting
time of an upper threshold $H>0$ until time $T>0$ for $i,k\in \mathbb N$
$$
z_{ik} = \E{\int_0^{\tau}t^i X_t^k\,dt}\leq\E{\int_0^T t^i X_t^k\,dt}\leq H^k\int_0^T t^i\,dt
=\frac{T^{i+1}H^k}{i+1}.$$
Thus, we fix a scaling vector $\vec d$ with entries $d_{ik}={T^{i+1}H^k}$ in
the same order
as the canonical base vector~\eqref{eq:canoncial_basis}.
Using this scaling vector, we can define a scaling
matrix $D={\vec d}{\vec{d}}^{\T}$.
Clearly, $D \succeq 0$.
Now we can formulate the optimization~\eqref{eq:sdp_for_fpt}
over a scaled version $D^{-1}M(\vec{z'})$ instead of $M(\vec{z'})$.
The moment matrices of the exit location probabilities are scaled in the same
way.
Alternatively, one can use approximations such as moment closures
or bounds obtained by lower-order relaxations or solve a sequence
of problems, incrementally increasing
the time-horizon, and adjust the scaling accordingly~\cite{dowdy2018dynamic}.

\begin{figure}[ht]
    \centering
    \includegraphics[scale=0.65]{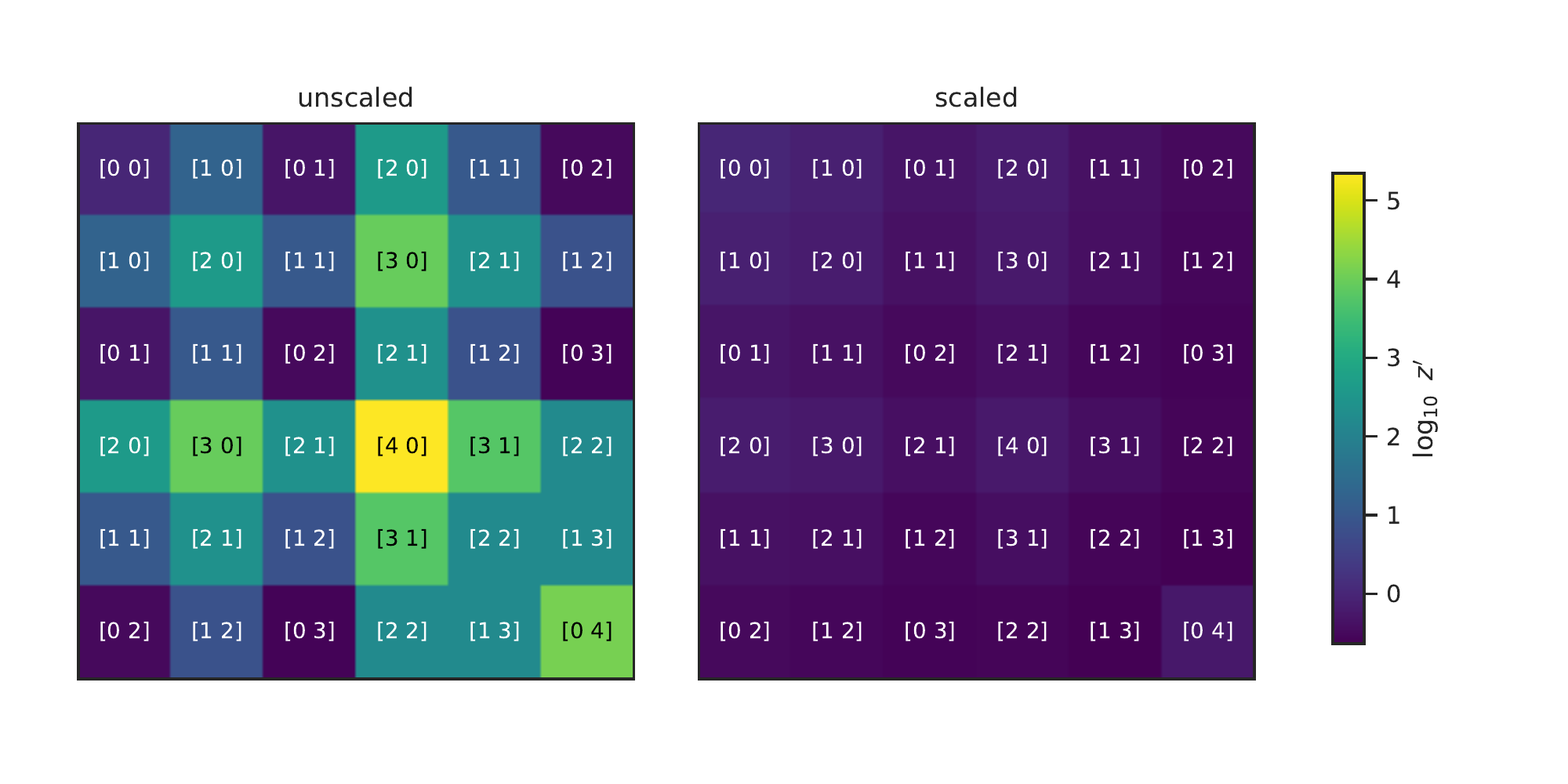}
    \vspace{-3em}
    \caption{The unscaled and scaled value the moment matrix proxy variable
    $M(\vec{z'})$ after optimization using MOSEK. The indices are given along the
    logarithmic (base 10) values. The unscaled version (left) shows large
    differences in magnitudes, while on the scaling suppresses
    these large variations (right). The case study used here is Model~\ref{model:dim},
    with a threshold $H=25$ for species $M$ and a time-horizon $T=1$. The
    relaxation order $r=2$. Therefore moments of orders up to $2r=4$ appear.}
    \label{fig:magnitudes}
\end{figure}

In Figure~\ref{fig:magnitudes} we illustrate the influence the scaling has on the
optimization variables. While the unscaled version shows large differences
between values, these differences become significantly smaller in the scaled version
of the problem.

\subsection{Case Studies}
We implemented and solved the SDP programs described above using optimization suite MOSEK~\cite{mosek} (version 9.1.2) via the CVXPY
interface~\cite{cvxpy} (version 1.0.24).

\subsubsection{Dimerization}
\begin{figure}[th]
    \centering
    \begin{minipage}{.49\textwidth}
    \centering
    \includegraphics[scale=.6]{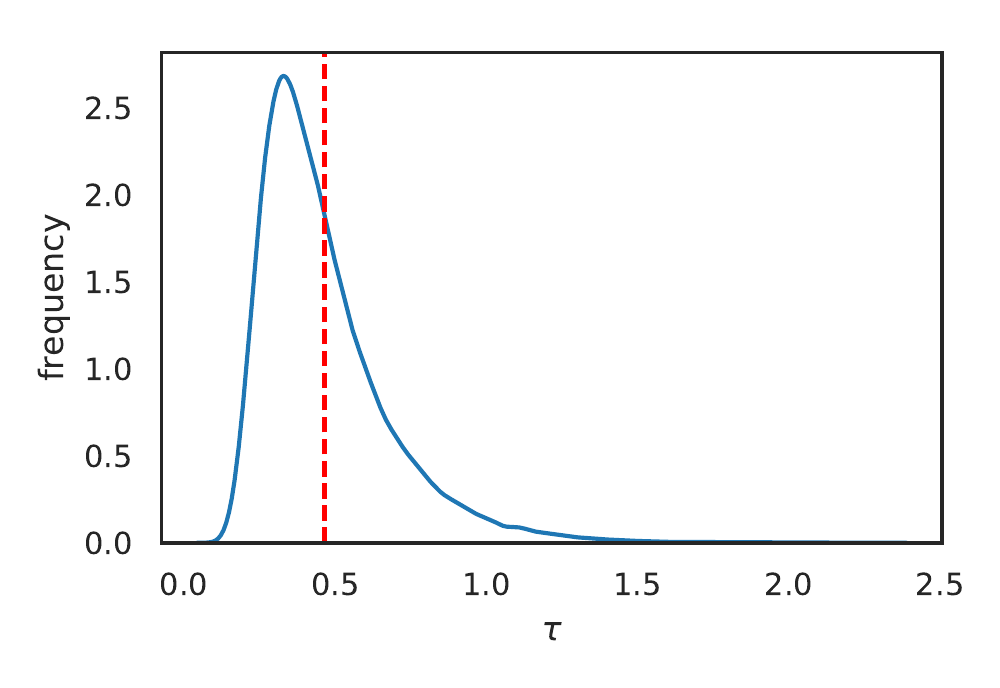}\\(a)
    \end{minipage}
    \begin{minipage}{.49\textwidth}
    \centering
    \includegraphics[scale=.6]{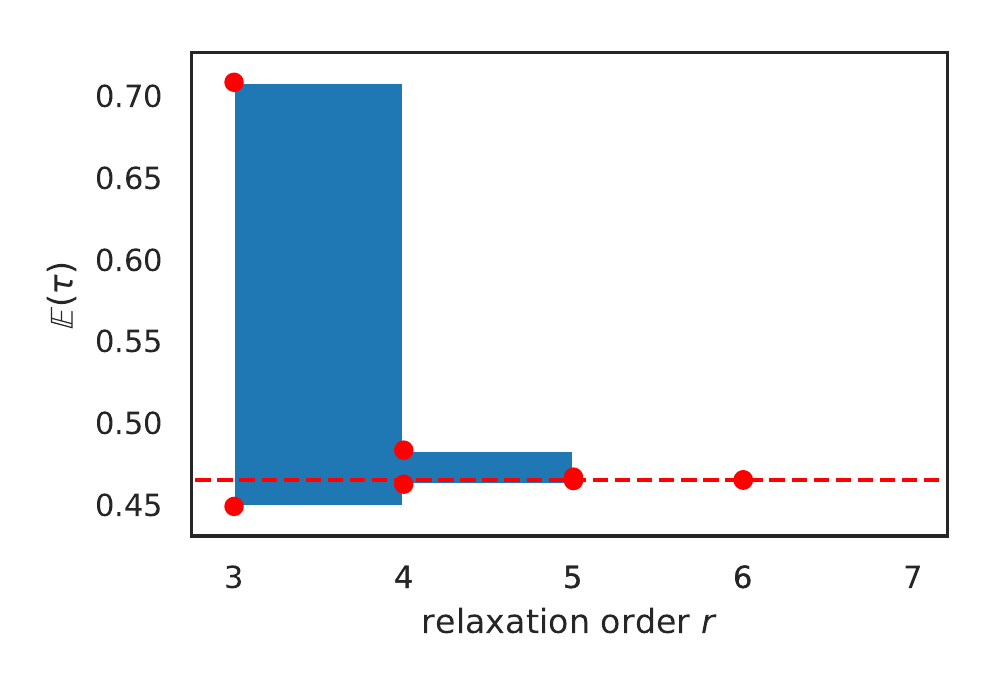}\\(b)
    \end{minipage}
    \caption{First passage times for Model 1 with $\tau=\inf\{t\geq 0\mid X_t \geq 10\}\land \infty$.
    The dashed red line denotes the sampled MFPT\@.
    (a) The distribution of $\tau$ estimated based on 100,000 SSA samples.
    (b) The bounds based on the SDP in~\eqref{eq:sdp_for_fpt} with different moment
    orders.\label{fig:dim_fpt}}
\end{figure}

As a first case study, we use Model~\ref{model:dim} with parameters $\lambda=100$ and $\delta=0.2$.
In this model, we are interested in the time at which the number of agents of type $M$
surpasses a threshold of 25 before some time-horizon $T$,
i.e.\ $\tau=\inf\{t\geq 0\mid X_t \geq 25\}\land T$.
First, we set no finite time-horizon $T$, i.e.\ $T=\infty$.
This is achieved by dropping the moments $\vec y_2$
of measure $\nu_2$ in the linear constraints~\eqref{eq:sdp_for_fpt}.
This can be done because the threshold on $M$ makes the state-space finite
and therefore the first passage time distribution is a phase-type distribution
which possesses finite moments \cite[Chapter 7.6]{stewart2009probability}.
The empirical FPT distribution based on 100,000 SSA simulations is given in Figure~\ref{fig:dim_fpt}a
and the bounds, given different moment orders, are given in Figure~\ref{fig:dim_fpt}b.
As we can see in Figure~\ref{fig:dim_fpt}b, the bounds capture the MFPT precisely for orders 5, 6.
The difference between upper and lower bound decreases roughly exponentially with increasing relaxation order $r$.
We found that this trend was consistent among the case studies presented here (cf.\ Figure~\ref{fig:convergence}).

\begin{figure}[t]
    \centering
    \begin{minipage}{.49\textwidth}
    \centering
    \includegraphics[scale=.6]{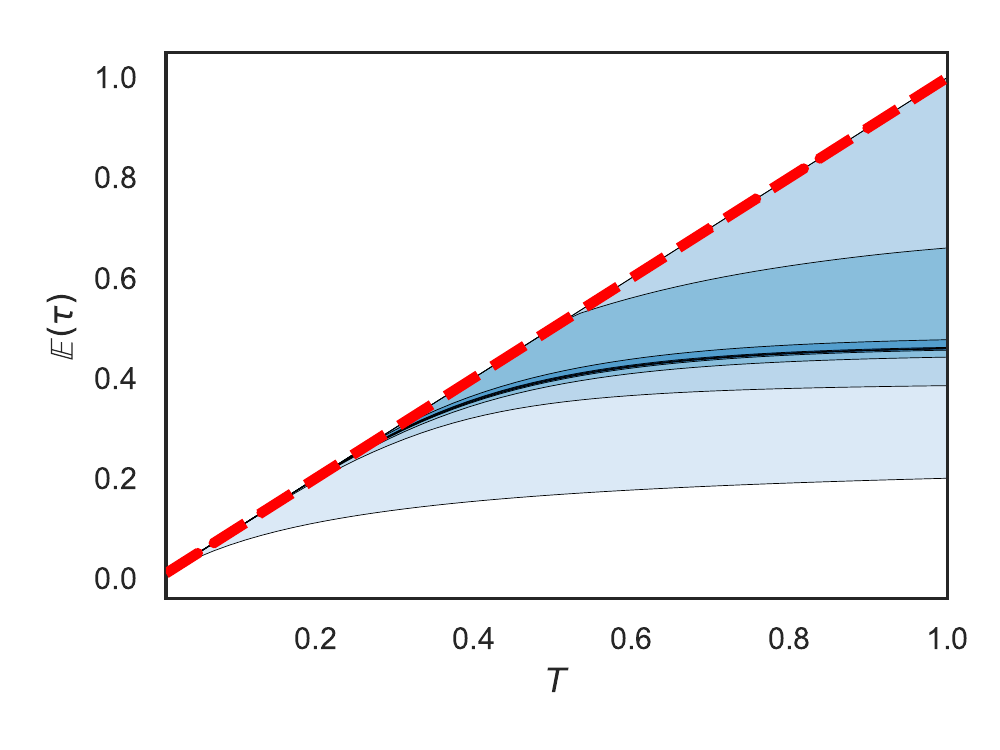}\\(a)
    \end{minipage}
    \begin{minipage}{.49\textwidth}
    \centering
    \includegraphics[scale=.6]{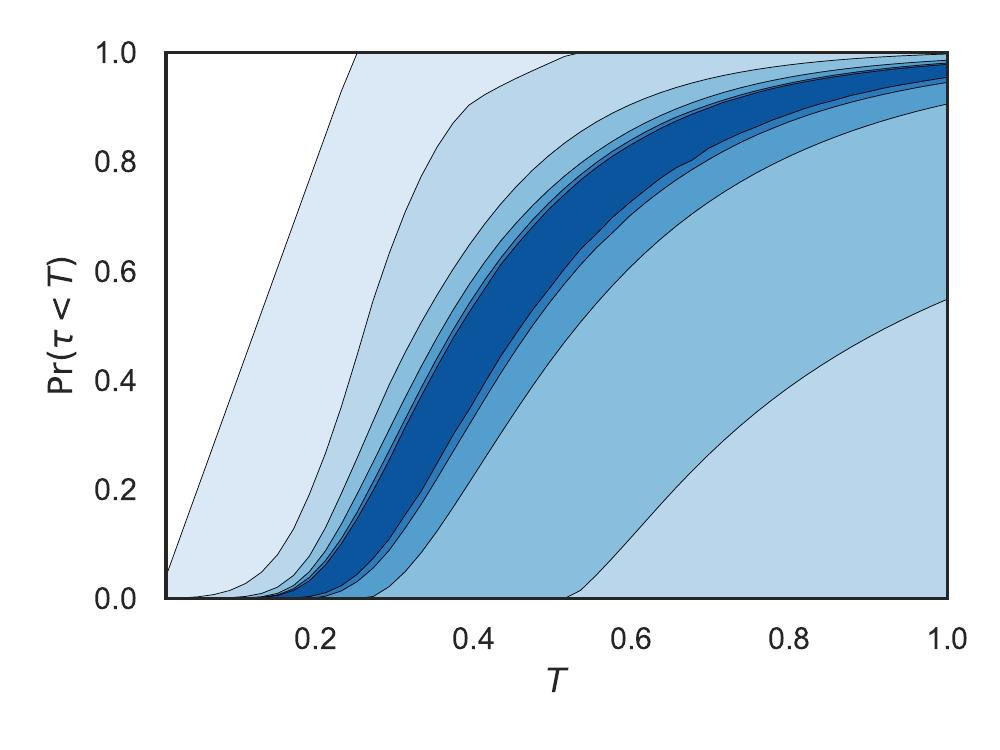}\\(b)
    \end{minipage}
    % \begin{minipage}{.15\textwidth}
    % legend
    % \end{minipage}
    \caption{First passage times for the dimerization model with
    $\tau=\inf\{t\geq 0\mid X_t \geq 25\}\land T$.
    The results for SDP relaxations of orders 1 (light) to 6 (dark) are shown.
    (a) The bounds on the MFPT for differing time-horizons $T$.
    (b) Bounds on the probability to reach the threshold before time $T$.\label{fig:dim_fpt_fin}}
\end{figure}

Next, we look at first passage times within a finite time-horizon $T$.
In Figure~\ref{fig:dim_fpt_fin}a we summarize the bounds obtained for the MFPT over $T$.
While low-order relaxations (light) give rather loose bounds, the bounds are already fairly tight
when using $r=4$.
In many cases, hitting probabilities, that is, the probability of reaching the
threshold before time $T$, are of particular interest.
This is done by switching the optimization objective in~\eqref{eq:sdp_for_fpt} from the mass of the
expected occupation measure $\xi$ to the mass of $\nu_1$.
In terms of moments, the objective changes from $z_{00}$ to $y_{10}$.
The need for such a scenario often arises in the context of model checking, where one might be
interested in the probability of a population exceeding a critical threshold.
By varying the time-horizon, we are able to recover bounds on the cumulative density $F(t) = \Pr(X_s=H\mid s<t)$ of
the first passage time (Fig.~\ref{fig:dim_fpt_fin}b).

Finally, we look at turn to the dimer species $D$ that is synthesized by the combination of
two monomers $M$. Here, we look at the time until the agents of type $D$ exceed a threshold of five
with a time-horizon $T=1$. Note that we do not limit the number of $M$ agents. Therefore
the analyzed state-space is countably infinite. As in the previous two examples, we
observe a roughly exponential decrease in interval size with increasing relaxation
order $r$ (cf.\ Fig.~\ref{fig:convergence} and Table~\ref{tab:bounds}).

\subsubsection{Parallel Dimerizations}
As a second study, we consider a 2-dimensional model by combining two
independent dimerizations.
\begin{model}[Parallel independent dimerizations]\label{model:double_dim}
$$\varnothing\xrightarrow{10^4}M_1,\quad 2M_1\xrightarrow{0.1}D_1,\quad \varnothing\xrightarrow{10^4}M_2,\quad 2M_2\xrightarrow{0.1}D_2$$
\end{model}
As a FPT we consider the time at which either $M_1$ or $M_2$ surpasses a threshold of 200 or a time-horizon of $T=10$
is reached, i.e.
$$ \tau=\inf\{t\geq 0\mid X_t^{(M_1)} \geq 200\}\land \inf\{t\geq 0\mid X_t^{(M_2)} \geq 200\}\land 10\,.$$
As before, we ignore the product species $D_1$ and $D_2$ since they do not influence $\tau$.
% Still, the possible state-space reaches a size of $200^2=40,\!000$.
% \luca{40000 is very small... I would not emphasize it}\todo{yes, but the only reason for this case study ist that we have a ``larger'' relevant state-space}
The SSA   (using $n=10,\!000$ runs) gives the estimate $\E{\tau}\approx 0.028378$ %\pm 1.3025e-04$ (99\%-CI)
which is captured tightly by the SDP bounds (cf.\ Table~\ref{tab:bounds}).
For higher relaxation orders $r \geq 5$  numerical issues prevented the solution of the
corresponding SDPs.

\subsection{Hybrid Models and Multi-Modal Behavior}
The analysis of switching times is a particularly interesting case of FPTs that
arises in many   contexts.
Often mode switching in such systems can be described a modulating Markov process
whose switching rates may depend on the system state (e.g.\ the population sizes).
In biological applications, mode switching often describes a change of the
DNA state~\cite{hasenauer2014method,stekel2008strong} and the analysis of
switching time distribution is of particular interest~\cite{spieler2011model,barzel2008calculation}.
In the context of PCTMCs, the state-space of such models can be given as $$\mathcal{S}=
\mathbb{N}^{\tilde{n}_S}\times {\{0,1\}}^{\hat{n}_S}\,.$$
This state is modeled by  $\hat{n}_S$ population variables with
binary domains. Therefore, at each time point, the state of these modulator variables
is given by a set of Bernoulli random variables.
When considering the moments of
such a variable $X$, clearly $\E{X^m}=\E{X}=\Pr(X=1)$ for all $m\geq 1$.

% We can use this fact two ways: First we could use the same moment
% constraints
% as above and impose additional equality constraints on the moments matrices
% to ensure $\E{X^m}=\E{X}$, $m\geq 1$.

We apply a  split of variables $\vec X_t$  into the high count part ${\vec{\tilde{X}}}_t$ and the binary
part ${\vec{\hat{X}}}_t$ to
the expectations in~\eqref{eq:mom_ode}. Similarly, we split   $\vec v_j$ and
with a case
distinction over the mode variable,
we arrive at a similar result as in~\cite{hasenauer2014method}:
\begin{equation}\label{eq:mcm}
\begin{split}
    \frac{d}{dt}\E{\vec{\tilde{X}}^{{\vec m}}_t 1_{=\vec y}({\vec{\hat{X}}}_t)}
   % =&\sum_{j=1}^{n_R}
   % \E{\left(\left(\vec{\tilde{X}}_t+\vec{\tilde{v}}_j\right)^{\vec
   % m}1_{=\vec y}({\vec{\hat{X}}}_t+\vec{\hat{v}}_j)
   %     -{\vec{\tilde{X}}}_t^{\vec
   %     m}1_{=\vec y}({\vec{\hat{X}}}_t)\right)\alpha(\vec X_t) }\\
    =&\sum_{j=1}^{n_R}
        \E{{\left(
                {\vec{\tilde{{X}}}}_t+\vec{\tilde{{v}}}_j
            \right)}^{\vec{m}}\alpha_j(\vec{\tilde{{X}}}_t, \vec{{y}} -
            \vec{\hat{v}}_j)
            1_{={\vec{y}- \vec{\hat{v}}_j}}({\vec{\hat{X}}}_t )}\\
    &- \sum_{j=1}^{n_R}
    \E{{\vec{\tilde{{X}}}}_t^{\vec{m}}\alpha_j(\vec{\tilde{{X}}}_t,
                {\vec{y}})
            1_{=\vec y}({\vec{\hat{X}}}_t)}\,.
            \end{split}
\end{equation}
Similarly to the general moment case, we can derive a constraint, by multiplying with a time-weighting factor
and integrating.

For simplicity, here we assume $\tilde n_S=\hat n_S=1$.
Fixing appropriate sequences ${(c_i)}_i$, ${(m_i)}_i$, ${(k_i)}_i$, and
${(y_i)}_i$ the constraint has the following form.
\begin{equation}\label{eq:mcm_constraint}
    \begin{split}
    &\sum_{y\in\{0,1\}}{H}^{m}\E{\tau^k;{{\hat{X}}}_{\tau}= y, \tau
    < T}
    +
    T^k\E{{{\tilde{X}}}_T^{m};{{\hat{X}}}_T= y,\tau=T}\\
      =\; &  0^k{{\tilde{x}}}_0^{m} 1_{=  y}({\hat{x}}_0) +
    \sum_i c_i %\sum_{y\in\{0,1\}}
        \E{\int_0^{\tau}t^{k_i}{{{\tilde{X}}}_t}^{{m}_i}\,dt;
            {{\hat{X}}}_t= y_i}\\
    \end{split}
\end{equation}
This way we can decompose the moment matrices such that for each mode $y\in\{0,1\}$,
we have moment matrices composed of the respective partial moments.
To this end, let $z^{(y)}_m$ be the partial moment w.r.t.\ ${{\hat{X}}}= y$.
The moment constraint over the partial moments has a linear structure:
\begin{equation}
0=y_{1k} {H}^{m} - y_{2m}T^k - 0^k x_0^{m}
+\sum_i c_i z^{(y_i)}_{k_i  m_i}\,.
\end{equation}

\begin{table}[t]
\centering
    \caption{MFPT bounds on Models~\ref{model:dim}, ~\ref{model:double_dim}, and~\ref{model:gexpr}.\label{tab:bounds}}
\begin{tabular}{l@{\hspace{2em}}l@{\hspace{2em}}r@{\hspace{2ex}}r@{\hspace{2ex}}r@{\hspace{2ex}}r@{\hspace{2ex}}r}
    \toprule
    Model & & \multicolumn{5}{c}{Relaxation Order $r$}\\
        \cmidrule{3-7}
        & & 1 & 2        & 3        & 4       & 5       \\
        \midrule
        Dimerization (Model~\ref{model:dim}) & lower & 0.0909 & 0.2661 & 0.2845 & 0.2867 & 0.2871\\
        $X_t^{(D)}\geq 5,\;\; T= 1$ & upper & 1.0000 & 0.3068 & 0.2932 & 0.2886 & 0.2875  \\
         \midrule
         Double Dim. (Model~\ref{model:double_dim}) & lower &0.0010 & 0.0250 & 0.0275 & 0.0280 & 0.0280\\
         & upper & 10.0000 & 0.0575 & 0.0323 & 0.0299 & 0.0290 \\
         \midrule
         Gene Expression (Model~\ref{model:gexpr}) & lower & 4.0000 & 6.0028 & 6.2207 & 6.3377 & 6.3772  \\
         & upper & 10.7179 & 6.4619 & 6.4079 & 6.4004 & 6.3835 \\\bottomrule
\end{tabular}
\end{table}

\subsubsection{Gene Expression with Negative Feedback}
As an instance of a multi-modal system, we consider a simple gene expression with self-regulating
negative feedback which is a common pattern in many genetic circuits~\cite{stekel2008strong}.

\begin{model}[Negative self-regulated gene expression]\label{model:gexpr}
This model consists of a gene state that is either on or off, i.e.\ $X^{D_{\text{on}}}_t
+X^{D_{\text{off}}}_t = 1$, $\forall t\geq 0$. Therefore the system has two \emph{modes}.
$$
D_{\text{on}} \xrightarrow{\tau_{0}} D_{\text{off}}, \quad
D_{\text{off}} \xrightarrow{\tau_{1}} D_{\text{on}}, \quad
D_{\text{on}} \xrightarrow{\rho} D_{\text{on}} + P, \quad
$$
$$
P\xrightarrow{\delta}\varnothing,\quad
P + D_{\text{on}} \xrightarrow{\gamma} D_{\text{off}}
$$
The model parameters are $(\tau_0,\tau_1,\rho,\delta,\gamma)=(10,10,2,0.1,0.1)$ and
$X_0^{(D_{\text{off}})}=1$, $X_0^{(P)}=0$ a.s.
\end{model}

As a first passage time we consider $$\tau=\inf\{t\geq 0\mid X_t^{(P)} \geq
5\}\land 20\,.$$

The results are summarized in Table~\ref{tab:bounds}.
The estimated MFPT based on $100,\!000$
SSA samples is $\E{\tau}\approx 6.37795\pm0.02847$ at $99\%$ confidence level.
Note that our SDP solution for $r=5$ yields tighter moment bounds than
the statistical estimation.% based on $100,\!000$ trajectories.

\begin{figure}[t]
    \centering
    \includegraphics[scale=0.5]{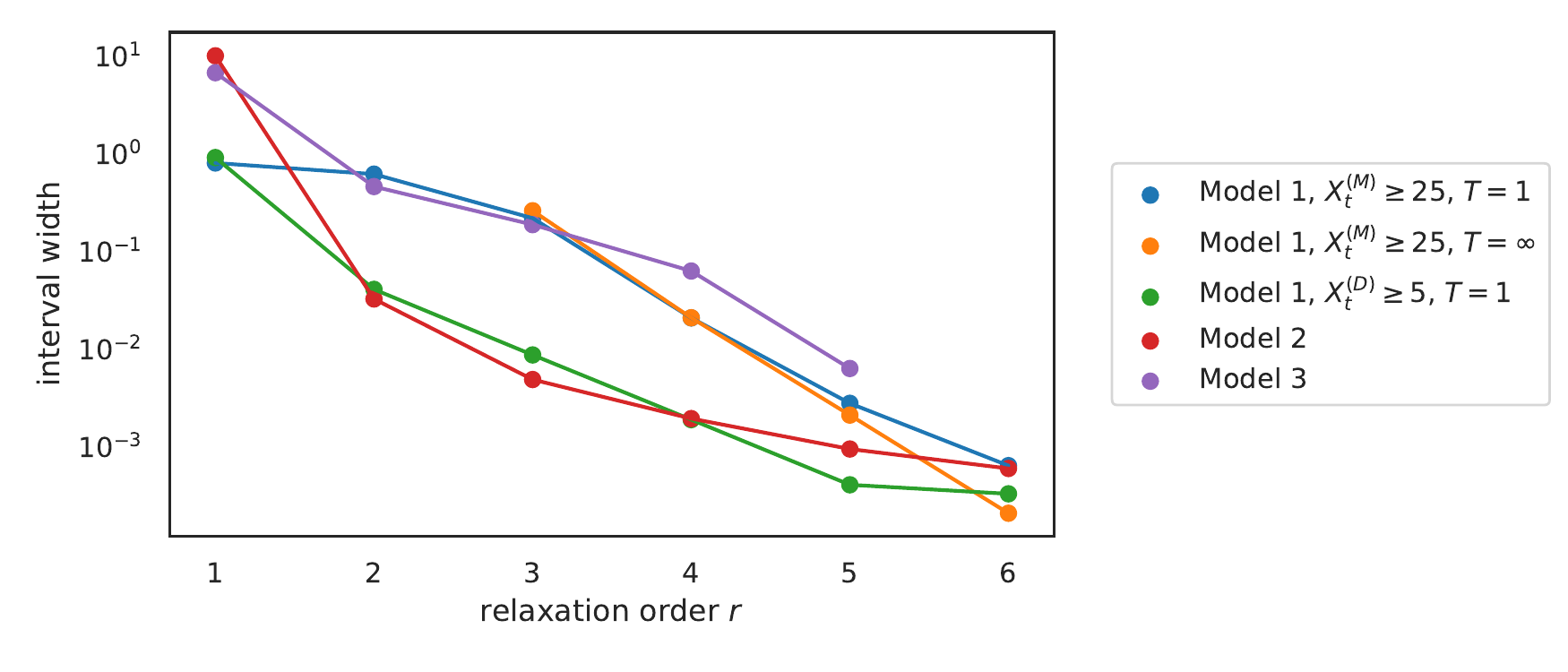}
    \caption{The interval width, i.e.\ the difference between upper and lower bound,
    for different case studies and targeted first passage times against the order $r$
    of the SDP relaxation.\label{fig:convergence}}
\end{figure}

In Fig.~\ref{fig:convergence} we summarize our results about the decrease of the interval widths for increasing relaxation order $r$ by plotting them on a log-scale.
We see an approximately exponential decrease with increasing $r$.
The semi-definite programs above were all solved within at most a few seconds.

\section{Conclusion}\label{sec:conclusion}
Numerical methods to compute reachability probabilities and first passage times
for continuous-time Markov chains that are based on an exhaustive exploration of the state-space
are exact up to numerical precision. Such methods, however, do not scale and cannot be efficiently applied to models with large or infinite state-spaces,
an issue exacerbated in population models. Moment-based methods offer an alternative analysis approach
for PCTMCs, which scales with the number of different populations in the system
but are approximations with little or no control of the error.
In this paper,
we bridge this gap by proposing a rigorous approach to derive bounds on first passage
times and reachability probabilities, leveraging a semi-definite programming formulation
based on appropriate moment constraints. 

The method we propose is shown to be accurate in several examples. It does, however, suffer, like all moment-based methods, from numerical instabilities in the SDP solver, caused by the fact that moments typically span several orders of magnitude. We proposed a scaling of moments to mitigate this effect. 
However, the scaling only addresses the moment matrices but not the linear constraints
which still contain values with varying orders of magnitudes.
Therefore, we plan as future work to investigate an appropriate scaling for the linear constraints
or to redefine the moment
constraints (e.g.\ using an exponential time weighting~\cite{dowdy2018dynamic}).
Based on this investigation, we expect to make this approach applicable to
more problems including, for example, the computation of bounds of rare event probabilities.
%Numerical instabilities due to moment values of largely differing orders of magnitudes are a current limitation of all moment-based methods.
We also expect that the development of more sophisticated scaling techniques will  improve approximate moment-based methods.

Furthermore, moment-based analysis approaches have shown to be successful
in a wide range of applications such as optimal control problems or the estimation
of densities~\cite{lasserre2010moments}.
We expect that our proposed ideas   can be adapted to a wider range
of stochastic models such as stochastic hybrid systems, exhibiting partly deterministic dynamics.

\subsubsection{Acknowledgements} We would like to thank Andreas Karrenbauer for helpful comments on
the usage of SDP solvers and Gerrit Gro\ss{}mann for the valuable comments on this manuscript.
This work is supported by the DFG project ``MULTIMODE'',
and partially supported by the italian PRIN project ``SEDUCE'' n.\ 2017TWRCNB\@.

\bibliographystyle{splncs04}
\bibliography{paper.bib}

\begin{thebibliography}{10}
\providecommand{\url}[1]{\texttt{#1}}
\providecommand{\urlprefix}{URL }
\providecommand{\doi}[1]{https://doi.org/#1}

\bibitem{andreychenko2011parameter}
Andreychenko, A., Mikeev, L., Spieler, D., Wolf, V.: Parameter identification
  for {M}arkov models of biochemical reactions. In: International Conference on
  Computer Aided Verification. pp. 83--98. Springer (2011)

\bibitem{aziz1996verifying}
Aziz, A., Sanwal, K., Singhal, V., Brayton, R.: Verifying continuous time
  {M}arkov chains. In: International Conference on Computer Aided Verification.
  pp. 269--276. Springer (1996)

\bibitem{backenkohler2017moment}
Backenk{\"o}hler, M., Bortolussi, L., Wolf, V.: Moment-based parameter
  estimation for stochastic reaction networks in equilibrium. IEEE/ACM
  transactions on computational biology and bioinformatics  \textbf{15}(4),
  1180--1192 (2017)

\bibitem{backenkohler2019control}
Backenk{\"o}hler, M., Bortolussi, L., Wolf, V.: Control variates for stochastic
  simulation of chemical reaction networks. In: Bortolussi, L., Sanguinetti, G.
  (eds.) Computational Methods in Systems Biology. pp. 42--59. Springer, Cham
  (2019)

\bibitem{baier2003model}
Baier, C., Haverkort, B., Hermanns, H., Katoen, J.P.: Model-checking algorithms
  for continuous-time {M}arkov chains. IEEE Transactions on software
  engineering  \textbf{29}(6),  524--541 (2003)

\bibitem{baier2000model}
Baier, C., Haverkort, B., Hermanns, H., Katoen, J.P.: Model checking
  continuous-time {M}arkov chains by transient analysis. In: International
  Conference on Computer Aided Verification. pp. 358--372. Springer (2000)

\bibitem{barzel2008calculation}
Barzel, B., Biham, O.: Calculation of switching times in the genetic toggle
  switch and other bistable systems. Physical Review E  \textbf{78}(4),  041919
  (2008)

\bibitem{bel2009simplicity}
Bel, G., Munsky, B., Nemenman, I.: The simplicity of completion time
  distributions for common complex biochemical processes. Physical biology
  \textbf{7}(1),  016003 (2009)

\bibitem{bernardo2016}
Bernardo, M., De~Nicola, R., Hillston, J. (eds.): Formal {Methods} for the
  {Quantitative} {Evaluation} of {Collective} {Adaptive} {Systems}, Lecture
  {Notes} in {Computer} {Science}, vol.~9700. Springer International
  Publishing, Cham (2016)

\bibitem{bogomolov2015adaptive}
Bogomolov, S., Henzinger, T.A., Podelski, A., Ruess, J., Schilling, C.:
  Adaptive moment closure for parameter inference of biochemical reaction
  networks. In: International Conference on Computational Methods in Systems
  Biology. pp. 77--89. Springer (2015)

\bibitem{bortolussi2013}
Bortolussi, L., Hillston, J., Latella, D., Massink, M.: Continuous
  approximation of collective system behaviour: {A} tutorial. Performance
  Evaluation  \textbf{70}(5),  317--349 (May 2013)

\bibitem{bortolussi2013model}
Bortolussi, L., Lanciani, R.: Model checking {M}arkov population models by
  central limit approximation. In: International Conference on Quantitative
  Evaluation of Systems. pp. 123--138. Springer (2013)

\bibitem{bortolussi2014stochastic}
Bortolussi, L., Lanciani, R.: Stochastic approximation of global reachability
  probabilities of {M}arkov population models. In: Computer Performance
  Engineering - 11th European Workshop, {EPEW} 2014, Florence, Italy, September
  11-12, 2014. Proceedings. pp. 224--239 (2014)

\bibitem{chen2011time}
Chen, T., Diciolla, M., Kwiatkowska, M., Mereacre, A.: Time-bounded
  verification of {CTMCs} against real-time specifications. In: International
  Conference on Formal Modeling and Analysis of Timed Systems. pp. 26--42.
  Springer (2011)

\bibitem{chen2009quantitative}
Chen, T., Han, T., Katoen, J.P., Mereacre, A.: Quantitative model checking of
  continuous-time {M}arkov chains against timed automata specifications. In:
  2009 24th Annual IEEE Symposium on Logic In Computer Science. pp. 309--318.
  IEEE (2009)

\bibitem{david2015statistical}
David, A., Larsen, K.G., Legay, A., Miku{\v{c}}ionis, M., Poulsen, D.B.,
  Sedwards, S.: Statistical model checking for biological systems.
  International Journal on Software Tools for Technology Transfer
  \textbf{17}(3),  351--367 (2015)

\bibitem{dehnert2017storm}
Dehnert, C., Junges, S., Katoen, J.P., Volk, M.: A storm is coming: A modern
  probabilistic model checker. In: International Conference on Computer Aided
  Verification. pp. 592--600. Springer (2017)

\bibitem{cvxpy}
Diamond, S., Boyd, S.: {CVXPY}: A {P}ython-embedded modeling language for
  convex optimization. Journal of Machine Learning Research  \textbf{17}(83),
  ~1--5 (2016)

\bibitem{dowdy2018bounds}
Dowdy, G.R., Barton, P.I.: Bounds on stochastic chemical kinetic systems at
  steady state. The Journal of chemical physics  \textbf{148}(8),  084106
  (2018)

\bibitem{dowdy2018dynamic}
Dowdy, G.R., Barton, P.I.: Dynamic bounds on stochastic chemical kinetic
  systems using semidefinite programming. The Journal of chemical physics
  \textbf{149}(7),  074103 (2018)

\bibitem{engblom2006computing}
Engblom, S.: Computing the moments of high dimensional solutions of the master
  equation. Applied Mathematics and Computation  \textbf{180}(2),  498--515
  (2006)

\bibitem{gast2019}
Gast, N., Bortolussi, L., Tribastone, M.: Size expansions of mean field
  approximation: {Transient} and steady-state analysis. Performance Evaluation
  \textbf{129},  60 -- 80 (2019).
  \doi{https://doi.org/10.1016/j.peva.2018.09.005}

\bibitem{ghusinga2017exact}
Ghusinga, K.R., Vargas-Garcia, C.A., Lamperski, A., Singh, A.: Exact lower and
  upper bounds on stationary moments in stochastic biochemical systems.
  Physical biology  \textbf{14}(4),  04LT01 (2017)

\bibitem{gihmantheory}
Gihman, I., Skorohod, A.: The theory of stochastic processes ii. 1975

\bibitem{gillespie77}
Gillespie, D.: Exact stochastic simulation of coupled chemical reactions. The
  Journal of Physical Chemistry  \textbf{81}(25),  2340--2361 (1977)

\bibitem{gupta2014scalable}
Gupta, A., Briat, C., Khammash, M.: A scalable computational framework for
  establishing long-term behavior of stochastic reaction networks. PLoS Comput
  Biol  \textbf{10}(6),  e1003669 (2014)

\bibitem{hasenauer2014method}
Hasenauer, J., Wolf, V., Kazeroonian, A., Theis, F.J.: Method of conditional
  moments {(MCM)} for the chemical master equation. Journal of mathematical
  biology  \textbf{69}(3),  687--735 (2014)

\bibitem{hayden2012fluid}
Hayden, R.A., Stefanek, A., Bradley, J.T.: Fluid computation of passage-time
  distributions in large {M}arkov models. Theoretical Computer Science
  \textbf{413}(1),  106--141 (2012)

\bibitem{helmes2001computing}
Helmes, K., R{\"o}hl, S., Stockbridge, R.H.: Computing moments of the exit time
  distribution for {M}arkov processes by linear programming. Operations
  Research  \textbf{49}(4),  516--530 (2001)

\bibitem{hespanha2008moment}
Hespanha, J.: Moment closure for biochemical networks. In: 2008 3rd
  International Symposium on Communications, Control and Signal Processing. pp.
  142--147. IEEE (2008)

\bibitem{hinton2006prism}
Hinton, A., Kwiatkowska, M., Norman, G., Parker, D.: Prism: A tool for
  automatic verification of probabilistic systems. In: International Conference
  on Tools and Algorithms for the Construction and Analysis of Systems. pp.
  441--444. Springer (2006)

\bibitem{iyer2016first}
Iyer-Biswas, S., Zilman, A.: First-passage processes in cellular biology.
  Advances in Chemical Physics  \textbf{160},  261--306 (2016)

\bibitem{kashima2009polynomial}
Kashima, K., Kawai, R.: Polynomial programming approach to weak approximation
  of l{\'e}vy-driven stochastic differential equations with application to
  option pricing. In: 2009 ICCAS-SICE. pp. 3902--3907. IEEE (2009)

\bibitem{kazeroonian2014modeling}
Kazeroonian, A., Theis, F.J., Hasenauer, J.: Modeling of stochastic biological
  processes with non-polynomial propensities using non-central conditional
  moment equation. IFAC Proceedings Volumes  \textbf{47}(3),  1729--1735 (2014)

\bibitem{kuntz2017rigorous}
Kuntz, J., Thomas, P., Stan, G.B., Barahona, M.: Rigorous bounds on the
  stationary distributions of the chemical master equation via mathematical
  programming. arXiv preprint arXiv:1702.05468  (2017)

\bibitem{kuntz2018approximation}
Kuntz, J., Thomas, P., Stan, G.B., Barahona, M.: Approximation schemes for
  countably-infinite linear programs with moment bounds. arXiv preprint
  arXiv:1810.03658  (2018)

\bibitem{kuntz2019exit}
Kuntz, J., Thomas, P., Stan, G.B., Barahona, M.: The exit time finite state
  projection scheme: bounding exit distributions and occupation measures of
  continuous-time {M}arkov chains. SIAM Journal on Scientific Computing
  \textbf{41}(2),  A748--A769 (2019)

\bibitem{kwiatkowska2011prism}
Kwiatkowska, M., Norman, G., Parker, D.: Prism 4.0: Verification of
  probabilistic real-time systems. In: International conference on computer
  aided verification. pp. 585--591. Springer (2011)

\bibitem{lasserre2010moments}
Lasserre, J.B.: Moments, positive polynomials and their applications, vol.~1.
  World Scientific (2010)

\bibitem{lasserre2006pricing}
Lasserre, J.B., Prieto-Rumeau, T., Zervos, M.: Pricing a class of exotic
  options via moments and sdp relaxations. Mathematical Finance
  \textbf{16}(3),  469--494 (2006)

\bibitem{mikeev2013fly}
Mikeev, L., Neuh{\"a}u{\ss}er, M.R., Spieler, D., Wolf, V.: On-the-fly
  verification and optimization of dta-properties for large {M}arkov chains.
  Formal Methods in System Design  \textbf{43}(2),  313--337 (2013)

\bibitem{mosek}
{\relax MOSEK ApS}: MOSEK Optimizer API for C 8.1.0.67 (2018),
  \url{https://docs.mosek.com/8.1/capi/index.html}

\bibitem{munsky2009specificity}
Munsky, B., Nemenman, I., Bel, G.: Specificity and completion time
  distributions of biochemical processes. The Journal of chemical physics
  \textbf{131}(23),  12B616 (2009)

\bibitem{scs}
O'Donoghue, B., Chu, E., Parikh, N., Boyd, S.: {SCS}: Splitting conic solver,
  version 2.1.0. \url{https://github.com/cvxgrp/scs} (Nov 2017)

\bibitem{parrilo2003semidefinite}
Parrilo, P.A.: Semidefinite programming relaxations for semialgebraic problems.
  Mathematical programming  \textbf{96}(2),  293--320 (2003)

\bibitem{porter2016dynamical}
Porter, M.A., Gleeson, J.P.: Dynamical systems on networks. Frontiers in
  Applied Dynamical Systems: Reviews and Tutorials  \textbf{4} (2016)

\bibitem{sakurai2017convex}
Sakurai, Y., Hori, Y.: A convex approach to steady state moment analysis for
  stochastic chemical reactions. In: Decision and Control (CDC), 2017 IEEE 56th
  Annual Conference on. pp. 1206--1211. IEEE (2017)

\bibitem{sakurai2019bounding}
Sakurai, Y., Hori, Y.: Bounding transient moments of stochastic chemical
  reactions. IEEE Control Systems Letters  \textbf{3}(2),  290--295 (2019)

\bibitem{schnoerr2017efficient}
Schnoerr, D., Cseke, B., Grima, R., Sanguinetti, G.: Efficient low-order
  approximation of first-passage time distributions. Phys. Rev. Lett.
  \textbf{119},  210601 (Nov 2017). \doi{10.1103/PhysRevLett.119.210601}

\bibitem{schnoerr2015}
Schnoerr, D., Sanguinetti, G., Grima, R.: Comparison of different
  moment-closure approximations for stochastic chemical kinetics. The Journal
  of Chemical Physics  \textbf{143}(18),  185101 (Nov 2015).
  \doi{10.1063/1.4934990}

\bibitem{schnoerr2017survey}
Schnoerr, D., Sanguinetti, G., Grima, R.: Approximation and inference methods
  for stochastic biochemical kinetics—a tutorial review. Journal of Physics
  A: Mathematical and Theoretical  \textbf{50}(9),  093001 (Mar 2017).
  \doi{10.1088/1751-8121/aa54d9}

\bibitem{spieler2011model}
Spieler, D., Hahn, E.M., Zhang, L.: Model checking csl for {M}arkov population
  models. arXiv preprint arXiv:1111.4385  (2011)

\bibitem{stekel2008strong}
Stekel, D.J., Jenkins, D.J.: Strong negative self regulation of prokaryotic
  transcription factors increases the intrinsic noise of protein expression.
  BMC systems biology  \textbf{2}(1), ~6 (2008)

\bibitem{stewart2009probability}
Stewart, W.J.: Probability, Markov chains, queues, and simulation: the
  mathematical basis of performance modeling. Princeton university press (2009)

\bibitem{BuchWolkenhauer}
Ullah, M., Wolkenhauer, O.: Stochastic approaches for systems biology. Wiley
  interdisciplinary reviews. Systems biology and medicine  \textbf{2},  385--97
  (07 2009). \doi{10.1002/wsbm.78}

\bibitem{vandenberghe2010cvxopt}
Vandenberghe, L.: The cvxopt linear and quadratic cone program solvers. Online:
  http://cvxopt. org/documentation/coneprog. pdf  (2010)

\end{thebibliography}

\end{document}